\documentclass[aip,apl,numerical,reprint,linenumbers]{revtex4-1}
\usepackage{graphicx}%
\usepackage{bm}%
\usepackage{amssymb}
\usepackage[amssymb]{SIunits}
\usepackage[version=3]{mhchem}

\newcommand{\CNRSAddress}{CEA/CNRS joint team ``Nanophysics and Semiconductors'', Institut N\'eel-CNRS, BP 166, 25 rue des Martyrs, 38042 Grenoble Cedex 9, France}

\begin{document}

\title{Fabrication and tuning of plasmonic optical nanoantennas around droplet epitaxy quantum dots by cathodoluminescence}

\author{Gilles Nogues}
\email{gilles.nogues@grenoble.cnrs.fr}
\affiliation{\CNRSAddress}
\author{Quentin Merotto}
\affiliation{\CNRSAddress}
\author{Guillaume Bachelier}
\affiliation{``Near-field'' team, Institut N\'eel-CNRS, BP 166, 25 rue des Martyrs, 38042 Grenoble Cedex 9, France}
\author{Eun Hye Lee}
\affiliation{Center for Opto-Electronic Convergence Systems, Korea Institute of Science and Technology, Seoul 136-791, Korea}
\author{Jin Dong Song}
\affiliation{Center for Opto-Electronic Convergence Systems, Korea Institute of Science and Technology, Seoul 136-791, Korea}
\date{\today{}}


\begin{abstract}
We use cathodoluminescence to locate droplet epitaxy quantum dots with a precision $\lesssim$\unit{50}{\nano\meter} before fabricating nanoantennas in their vicinity by e-beam lithography. Cathooluminescence is further used to evidence the effect of the antennas as a function of their length on the light emitted by the dot. Experimental results are in good agreement with numerical simulations of the structures.
\end{abstract}

\maketitle

Ever decreasing sizes and low dimensionality of semiconducting heterostructures makes it possible to operate them in the quantum regime where single photons are produced and/or detected, paving the way to applications in quantum information processing and communication. In this context a key issue of nanooptics concerns the possibility to modify and control the properties of the light coming from a single solid-state emitter (direction, polarization, temporal profile). A potential route towards this goal is to embed it in a dielectric microcavity\cite{ENGLUNDVUCKOVIC_POINTDEFECT05} or a photonic waveguide\cite{ClaudonGerard_highlyefficientsingle-photon_10}. Another promising strategy is to extend the know-how of RF electrical engineering to the optical domain by placing metallic nanostructures in the vicinity of the emitter\cite{Greffet_Nanoantennaslightemission_05, MuhlschlegelPohl_ResonantOpticalAntennas_05}. They act as plasmonic optical nanoantennas able to enhance the emission rate of the emitter\cite{
doi:10.1021/nl902183y, SchietingerBenson_Plasmon-EnhancedSinglePhoton_09} and control its radiation pattern\cite{PhysRevLett.104.026802,Curto20082010,james:033106}. From a technological point of view, it is paramount to control both the emitter--antenna distance  with a precision $\lesssim$\unit{10}{\nano\meter}, and the caracteristic frequency of the plasmonic mode to tune the antenna to the emission wavelength. Precise positionning can be achieved with the help of AFM manipulation\cite{SchietingerBenson_Plasmon-EnhancedSinglePhoton_09, doi:10.1021/nl102548t} or chemical functionalization\cite{Curto20082010}, while standard top-down fabrication techniques are better suited to define the shape of the metal nanostructure and hence its resonant frequency. In this letter, we demonstrate  that cathodoluminescence (CL) combined to standard e-beam lithography is a powerfull tool to achieve both goals. We use this technique to locate droplet epitaxy quantum dots (QDs) before fabricating nanoantennas of different 
length around them. CL is further used to characterize the effect of the antennas on the QDs' emission properties as a function of the length.

\begin{figure}
 \centering
 \includegraphics[width=\columnwidth]{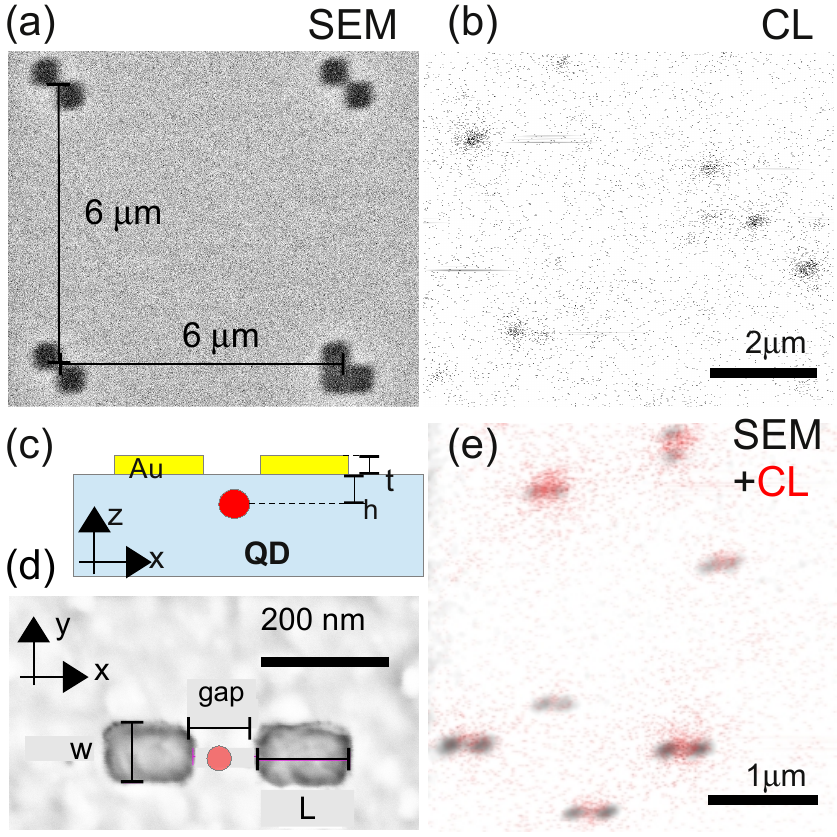}
  \caption{Simultaneously recorded images of (a) the SEM signal of the Au alignment marks deposited on the sample and (b) the CL signal of the droplet epitaxy quantum dots ($\lambda_{em}$=\unit{750}{\nano\meter}). (c) Cut and (d) top  field effect SEM microphotograph of the target QD/nanoantenna assembly; $h$=\unit{50}{\nano\meter}, gap \unit{100}{\nano\meter}, $w$=\unit{100}{\nano\meter}, $t$= \unit{35}{\nano\meter}, $L$ varies from 80 to \unit{350}{\nano\meter}. (e) Composite image of the fabricated Au nanoantennas observed in SEM (black channel) and of the QD spots (red channel).
  }\label{fig:fig1}
\end{figure}

Our target structure is shown in Fig.~\ref{fig:fig1}(c-d). It has already been shown to enhance the fluorescence of an ensemble of dye molecules\cite{doi:10.1021/nl0715847}. In our case a single \ce{GaAs}/\ce{AlGaAs} quantum dot is buried under the gap between two Au nanostrips.  QDs are produced using the modified droplet epitaxy method\cite{Kim20106500}. Atomic force microscopy measurements show a density of emitters of about \unit{3}{\micro\meter\rpsquared}. Each dot has a typical \unit{20}{\nano\meter} diameter and a height of \unit{10}{\nano\meter}. Capping layers of different thickness $h$ from 22 to \unit{70}{\nano\meter} are overgrown on top of the QD layer. Thermal annealing is necessary to activate the luminescence of the dots\cite{PSSR:PSSR201206369}. Ensemble photoluminescence measurements show that QDs emit between 750 and \unit{800}{\nano\meter}. Microphotoluminescence and cathodoluminescence allow us to observe that typical signal intensity from individual QD is dramatically reduced for 
thicknesses $h \le $\unit{40}{\nano\meter}. We attribute this observation to additional nonradiative losses generated by surface induced recombinations. Nanoantennas are fabricated on a sample with $h=$\unit{48}{\nano\meter} in order to ensure the smallest QD/antenna distance without having extra non-radiative losses. Before localization of individual quantum dots, \ce{Au} alignment marks are fabricated onto the substrate by e-beam lithography and sputtering of a Au layer followed by lift-off [Fig.~\ref{fig:fig1}(a)]. Marks are placed on a regular square array of period \unit{6}{\micro\meter}. 

The sample is then observed in CL at \unit{5}{\kelvin}. Light emitted by the excited dots is collected by a parabolic mirror and analyzed by a spectrometer. It is then detected by an avalanche photodiode (APD) at the output of the spectrometer. A standard SEM image of the marks [Fig.~\ref{fig:fig1}(a)] is simultaneously recorded with the CL image at emission wavelength $\lambda_{em}$=\unit{750}{\nano\meter} [Fig.~\ref{fig:fig1}(b)]. We limit the image acquisition time to a few seconds in order to limit the effect of thermal drift, surface contamination and charging of the substrate in the vicinity of the quantum dot. The presence of a QD is hence revealed by a cloud of single photon detection counts from the APD (black pixels on the image, typical cloud diameter $\sim$\unit{400}{\nano\meter}). The centers of the marks are determined by an edge detection algorithm on the SEM image. The CL signal intensity of each QD is obtained by summing all its corresponding pixels and its position is determined by their 
centroid. Combination of those informations yields the absolute coordinates of each dot with respect to the alignment mark array. 

Nanoantennas are fabricated by e-beam lithography on a \unit{100}{\nano\meter}-thick PMMA layer followed by sputtering of a \unit{35}{\nano\meter} thick layer of \ce{Au} and lift-off. Before exposing antennas the SEM e-beam is aligned by observing neighbouring marks. For all antennas the gap between the strips is \unit{100}{\nano\meter}, the width $w=$\unit{100}{\nano\meter} and the strips have a variable length $L$ ranging from 80 to                                                                                                                           \unit{360}{\nano\meter}.  

\begin{figure}
\begin{center}
  \includegraphics[width=\columnwidth]{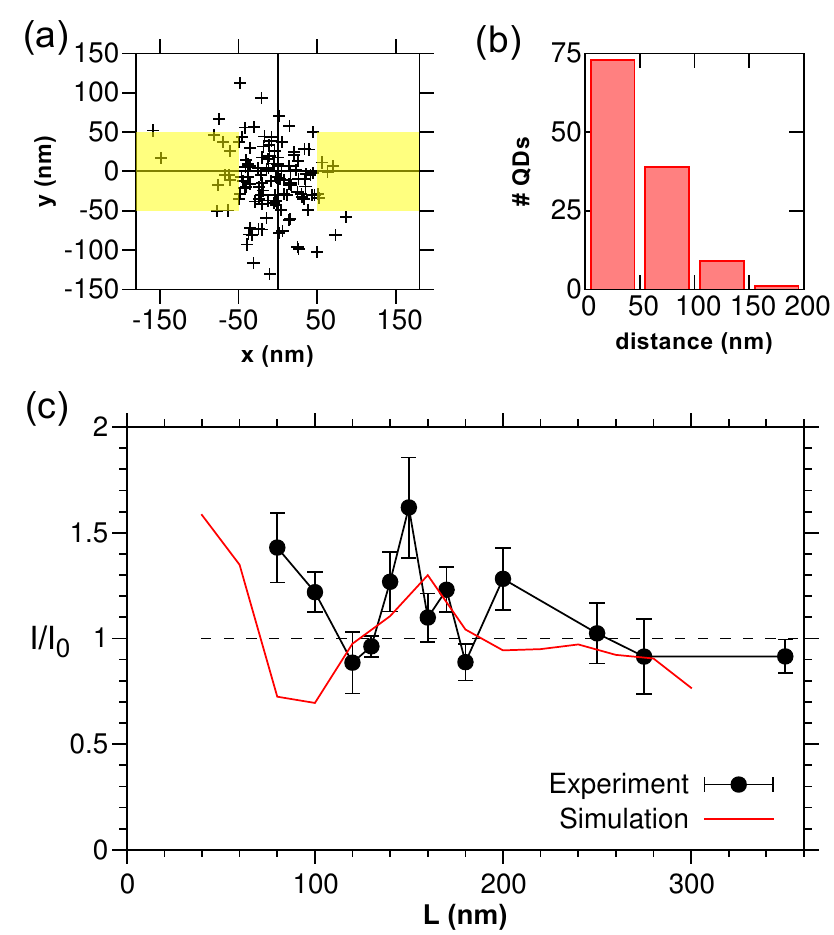}
\end{center}
  \caption{(a) Scatter plot of the QD positions measured by CL with respect to the nanoantenna center. The position of the strips is shown by yellow rectangles. (b) Histogram of QDs'distance distribution from antenna center. (c) Luminescence enhancement ratio $I/I_0$ as a function of antenna length $L$. Experimental points are compared to numerical simulations assuming a random distribution of dipole orientations}
  \label{fig:fig2}
\end{figure}

After fabrication of the nanoantennas, a second CL experiment is undertaken. Figure \ref{fig:fig1}(e) presents a composite image of the SEM signal (black channel) and CL signal (red channel). It clearly shows that each antenna coincides with a QD. In order to assess the performances of our fabrication process, we have fabricated 123 QD/antenna assemblies. For each system we measure the QD's position by its centroid as well as the antenna gap center on the SEM image. Figure \ref{fig:fig2}(a) shows a scatter plot of the relative positions of the QDs with respect to the nanoantennas and figure \ref{fig:fig2}(b) displays an histogram of their distance distribution. 70\% of the dots are within the \unit{100}{\nano\meter}$\times$\unit{100}{\nano\meter} square defined by the gap between the two nanostrips. Less than 8\% of them are at distances larger than \unit{100}{\nano\meter}. A qualitative study of those poorly fabricated assemblies shows that they correspond to systems located close to the edges of the images 
or to 
other dots. The centroid algorithm does not work properly in those conditions. The average distance to the antenna center is \unit{48}{\nano\meter}. Assuming that it is the sum of three independant equal sources of error (first localization, e-beam realignment before fabrication and second localization), one infers a typical localization error of \unit{28}{\nano\meter}. Our present error is larger than the target precision for fabricating state of the art  plasmonic structures, but it is nonetheless good enough to evidence coupling of the QD to the antenna in the case of a \unit{100}{\nano\meter} gap as is our case.

\begin{figure*}
  \begin{center}
    \includegraphics[width=\textwidth]{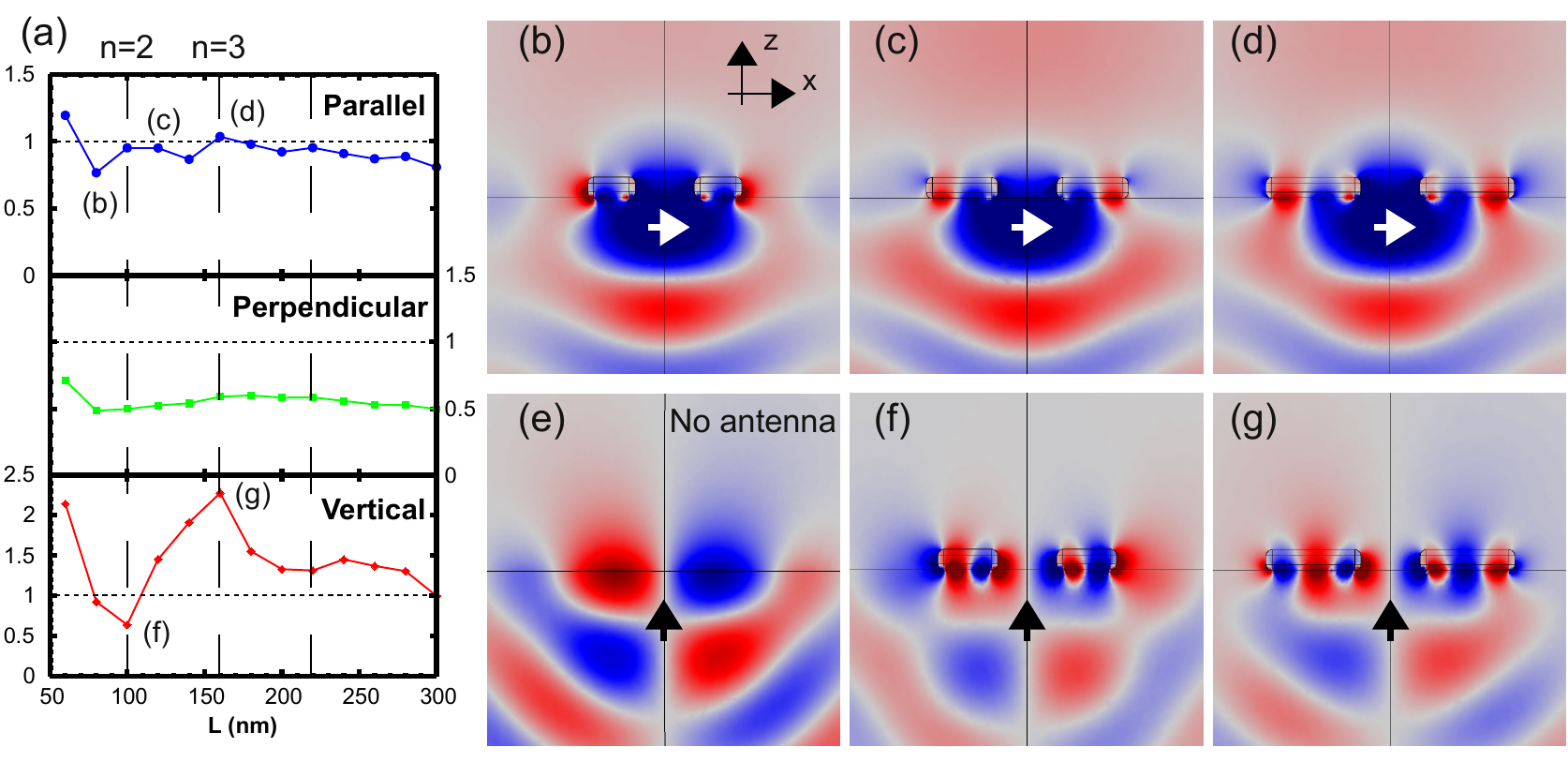}
    \caption{(a) simulated enhancement factors as a function of antenna length $L$ for a quantum dot corresponding to an oscillating dipole along the parallel ($x$), perpendicular ($y$) or vertical ($z$) direction. Maps of $Re[E_x]$ in the $(x,z)$ plane for a parallel dipole (b-d) and a vertical dipole (f-g) for different antenna lengths. The corresponding points are shown on figure (a) except for (e).}
    \label{fig:fig3}
  \end{center}
\end{figure*}

In order to evidence the effect of the nanoantennas on the QDs, we compare their CL signal in presence of the antenna $I$ to the one measured before fabrication $I_0$, with the same parameters of beam current, dwell-time and magnification. Hence, the excitation rate $\gamma_{exc}$ is the same for the two experiments. Moreover it is low enough to ensure that the quantum dots are not saturated, i.e. $\gamma_{exc} \ll \gamma_{tot}$ where $\gamma_{tot}$ is the relaxation rate of the quantum dot. Between 6 and 27 QD/antenna assemblies have been fabricated for each antenna length $L$. Figure \ref{fig:fig2}(c) presents the average ratio $I/I_0$ as a function of $L$. The corresponding error bars represent the statistical standard error on this value. One observes an enhancement of the signal (about 50\%) below \unit{80}{\nano\meter} and in the 140-\unit{170}{\nano\meter} range. This proves that it is possible to tune the antenna length in order to optimize the emission of light towards the CL detector. In our setup, 
 the collection mirror is placed just above the sample and we can assume that all the light emitted on the air side of the sample is collected. The measured 
intensity $I$ is then\cite{AngerNovotny_06} $I= h c/\lambda_{em} \cdot \gamma_{exc} \eta$, where $\eta$ is the quantum efficiency $\eta = \gamma_{r,\ air}/\gamma_{tot}=\gamma_{r,\ air}/(\gamma_{r,\ air}+\gamma_{r,\ sub}+\gamma_{nr})$. $\gamma_{nr}$ is the non-radiative relaxation rate of the quantum dot. The radiative relaxation rate $\gamma_r$ is the sum of the light emitted on the air-side and on the substrate side of the sample $\gamma_r=\gamma_{r,\ air}+\gamma_{r,\ sub}$.

A deeper understanding of our results arises from numerical simulations of the fabricated QD/antenna system with the finite-element software COMSOL. The QD is replaced by an electric dipole oscillating at frequency $\nu_{em}=c/\lambda_{em}$. We compute the power $P_{r,\ air}$ (resp. $P_{r,\ sub}$) radiated in the far-field towards the air side (resp. the substrate side) of the sample. We also evaluate the power $P_{abs}$ absorbed in the metallic nanoantennas due to the imaginary part of the index of refraction of Au at $\lambda_{em}$\cite{JohnsonChristy_OpticalConstantsof_72}. As we have previously checked that surface induced nonradiative recombinations are negligible for $h=$\unit{50}{\nano\meter}, we assume that all nonradiative losses are caused by absorption, i.e. $\gamma_{nr}\propto P_{abs}$. We also have $\gamma_{r,\ air}\propto P_{r,\ air}$ and $\gamma_{r,\ sub}\propto P_{r,\ sub}$. It is therefore possible to evaluate the quantum efficiency before and after the antenna fabrication. We perform theses 
calculations for fixed values of $w=$\unit{100}{\nano\meter}, $t=$\unit{35}{\nano\meter}, gap=\unit{100}{\nano\meter} and with various conditions of strip length $L$ or dipole orientation. We also plot the maps of the corresponding electromagnetic fields.

The simulations show that the dipole couples to antenna modes exhibiting 1 to 3 antinodes in each nanostrip along the $x$ axis (figure~\ref{fig:fig3}). From one mode to the next one, $L$ is increased by \unit{60}{\nano\meter}. This is in good agreement with half the expected surface plasmon wavelength $\lambda_p$ for the \ce{Au}/\ce{GaAs} interface at frequency $\nu_{em}$~\cite{PhysRevB.27.985, *[{see Equation (2.14) from }][{, dielectric constants are taken from Refs. 15 and 16}] MAierPLasmobook}. For a dipole oriented along the $y$ axis, the coupling to the antenna mode is poor. Hence the quantum efficiency is essentially flat and degraded to about 0.5 due to extra absorption in the metal [Fig.~\ref{fig:fig3}(a)]. In the case of a dipole along the $x$ axis, the excited antenna mode radiates either in phase with the dipole, resulting in an increase of the field received by the detector [Fig.~\ref{fig:fig3}(b)], or in phase opposition, leading to the reverse effect[Fig.~\ref{fig:fig3}(c)]. The relative phase 
of the plasmon oscillation is shifted by $\pi$ as $L$ spans across the resonance condition. As a consequence the luminescence 
enhancement is modulated with a period $\lambda_p/2$ with peaks at $L=100,160$ and \unit{220}{\nano\meter}. The effect of the antenna is dramatic in the case of a dipole along the $z$ axis, which does not radiate efficiently towards the detector in absence of antenna [Fig.~\ref{fig:fig3}(e)]. In this configuration the dipole excites antenna modes which efficiently radiate on the air side if they have an odd number of antinodes [Fig.~\ref{fig:fig3}(g)]. For even number of antinodes [Fig.~\ref{fig:fig3}(f)], the contributions of each antinode cancel in the far field and absorption dominates. As a result one observes a modulation of the emitted intensity versus $L$ with a period $\lambda_p$.

As opposed to stransky-Krastanov quantum dots, the orientation of the electric dipole in droplet epitaxy dots is not strongly constrained due to the absence of internal stress, relative large size and smooth interface. The comparison of the experimental results of figure~\ref{fig:fig2}(c) with the simulations of figure~\ref{fig:fig3}(a) suggests that a large fraction of systems have a dipole vertically aligned. The simulation data plotted on figure~\ref{fig:fig2}(c) correspond to an average of the three orientations. It is in good agreement with the experiment.

In summary, we have demonstrated the controlled coupling and tuning of Au nanoantennas to droplet epitaxy QDs using cathodoluminescence and standard electron-beam lithography. Our method offers the advantage of being spectrally selective and has a higher throughput than AFM nanomanipulation techniques. The enhancement factor of luminescence which we observe is well explained by numerical simulations. It could be dramatically increased with a smaller QD/antenna distance. We plan to improve the QD signal while reducing the capping layer. We have also developped a method for directly aligning the e-beam lithography setup on the CL signal, thus suppressing the realignement step onto ancillary marks\cite{DonatiniCLfabrication2010}. With those two improvements, we expect to fabricate coupled QD/antennas with typical depth and antenna gap of the order of \unit{20}{\nano\meter}. It is also possible to fabricate more complicated structures like clusters of nanoantennas is order to optimize the emission quantum 
efficiency\cite{doi:10.1021/nl0715847} and reach higher enhancement factors.

\begin{acknowledgements}
This research was supported by GRL "Development of innovative photonic devices" and LIA "Center for photonic research". The authors in KIST acknowledge the support from the KIST institutional program, including the Dream Project. We acknowledge the help of Institut N\'eel's technical support teams ``Nanofab'' (clean room) and ``optical engineering'' (CL setup, F. Donatini).    
\end{acknowledgements}


%

\end{document}